\documentclass[amsmath,amssymb,a4paper,preprint,showpacs,footinbib,floatfix,preprintnumbers,superscriptaddress]{revtex4-1}
\usepackage{graphicx,amsmath}
\usepackage[colorlinks=true,linkcolor=blue,citecolor=blue,urlcolor=blue]{hyperref}

\newcommand{\om}{\omega}
\newcommand{\ommax}{\om_{\rm max}}

 \newcommand{\rb}{\mathbf{r}}

\newcommand{\pdx}{\partial_x}
\newcommand{\pdt}{\partial_t}
\newcommand{\hp}{\phi}
\newcommand{\hpd}{\phi^*}

\begin{document}

\title{Instability of the superfluid flow as black-hole lasing effect}

\author{S. Finazzi}\email{stefano.finazzi@univ-paris-diderot.fr}
\affiliation{Laboratoire Mat\'eriaux et Ph\'enom\`enes Quantiques, Universit\'e Paris Diderot-Paris 7 and CNRS, B\^atiment Condorcet, 10 rue Alice Domon et L\'eonie Duquet, 75205 Paris Cedex 13, France.}
\author{F. Piazza}\email{francesco.piazza@ph.tum.de}
\affiliation{Technische Universit\"at M\"unchen, James-Franck-Stra{\ss}e 1, 85748 Garching, Germany}
\author{M. Abad}
\affiliation{INO-CNR BEC Center and Dipartimento di Fisica, Universit\`a di Trento, 38123 Povo, Italy}
\author{A. Smerzi}
\affiliation{QSTAR, INO-CNR and LENS, Largo Enrico Fermi 2, 50125, Firenze, Italy}
\author{A. Recati}\email{recati@science.unitn.it}
\affiliation{Technische Universit\"at M\"unchen, James-Franck-Stra{\ss}e 1, 85748 Garching, Germany}
\affiliation{INO-CNR BEC Center and Dipartimento di Fisica, Universit\`a di Trento, 38123 Povo, Italy}

\begin{abstract}
We show that the instability leading to the decay of the one-dimensional superfluid flow through a penetrable barrier are due to the black-hole lasing effect. This dynamical instability is triggered by modes resonating in an effective cavity formed by two horizons enclosing the barrier. The location of the horizons is set by $v(x)=c(x)$, with $v(x),c(x)$ being the local fluid velocity and sound speed, respectively. 
We compute the critical velocity analytically and show that it is univocally determined by the horizons configuration.
In the limit of broad barriers, the continuous spectrum at the origin of the Hawking-like radiation and of the Landau energetic instability is recovered.
\end{abstract}

\pacs{03.75.Lm, 03.75.Hh, 03.75.Kk -- XEC}

\maketitle

\section{Introduction} 

A very relevant and not fully understood problem in the field of supefluidity is the nature of the decay of the flow past a macroscopic obstacle. One the one hand, it is known \emph{how} superfluidity decays, namely via phase slippage induced by topological excitations like solitons or vortices \cite{anderson_1966}. However, on the other hand, the question \emph{why} superfluidity breaks down above a certain critical velocity has not yet found a conclusive answer, except from the limiting case where the obstacle is only a small perturbation of a homogeneous flow, in which case the Landau energetic instability takes place.

Due to its high degree of controllability, an optimal system for the investigation of superfluidity is a Bose-Einstein condensate (BEC) of ultracold dilute atoms \cite{pit_str_book}. In this system, obstacles are created by laser beams which can be precisely controlled over distances of the coherence length. This has allowed to observe superfluid decay and phase slippage with BECs both with moving obstacles in a bulk \cite{raman_1999, onofrio_2000,anderson_2010,dalibard_2012}, as well as with obstacles forming a constriction for the flow \cite{engels_2007,steinhauer_2007,nist_2011,hadzibabic_2012, nist_criticalvelocity_2013}. A quantitative theoretical description of BECs is provided by the Gross-Pitaevskii (GP) equation for the superfluid order parameter $\Psi(\rb,t)$. It contains the crucial ingredients giving rise to superfluidity - phase coherence and nonlinearity - and deals with a simple single classical field.

These same favorable features have made clear that BECs are also suitable to implement analog models of gravity \cite{visser_2001, beyer_2013}. Indeed, first experimental evidences of analog model phenomenology and/or quantum vacuum fluctuations have recently been reported \cite{steinhauer_2010, westbrook_2012, armijo_2012, chin_2013}. The analog model description not only puts the rich BEC phenomenology into a much broader context, but allows at the same time for new predictions and interpretations.

In this work, using concepts borrowed from analog models, we show that the supercurrent instability of the compressible flow through a constriction corresponds to the so called black-hole lasing effect\cite{cj,coutant,bhlaserBEC}. 
This provides on the one side a deeper understanding of a long standing problem in superfluidity and non-linear dynamics, on the other side it allows to introduce a new setup where the physics of sonic holes emerges naturally.
\begin{figure}[htp]
\includegraphics[width=12.0cm]{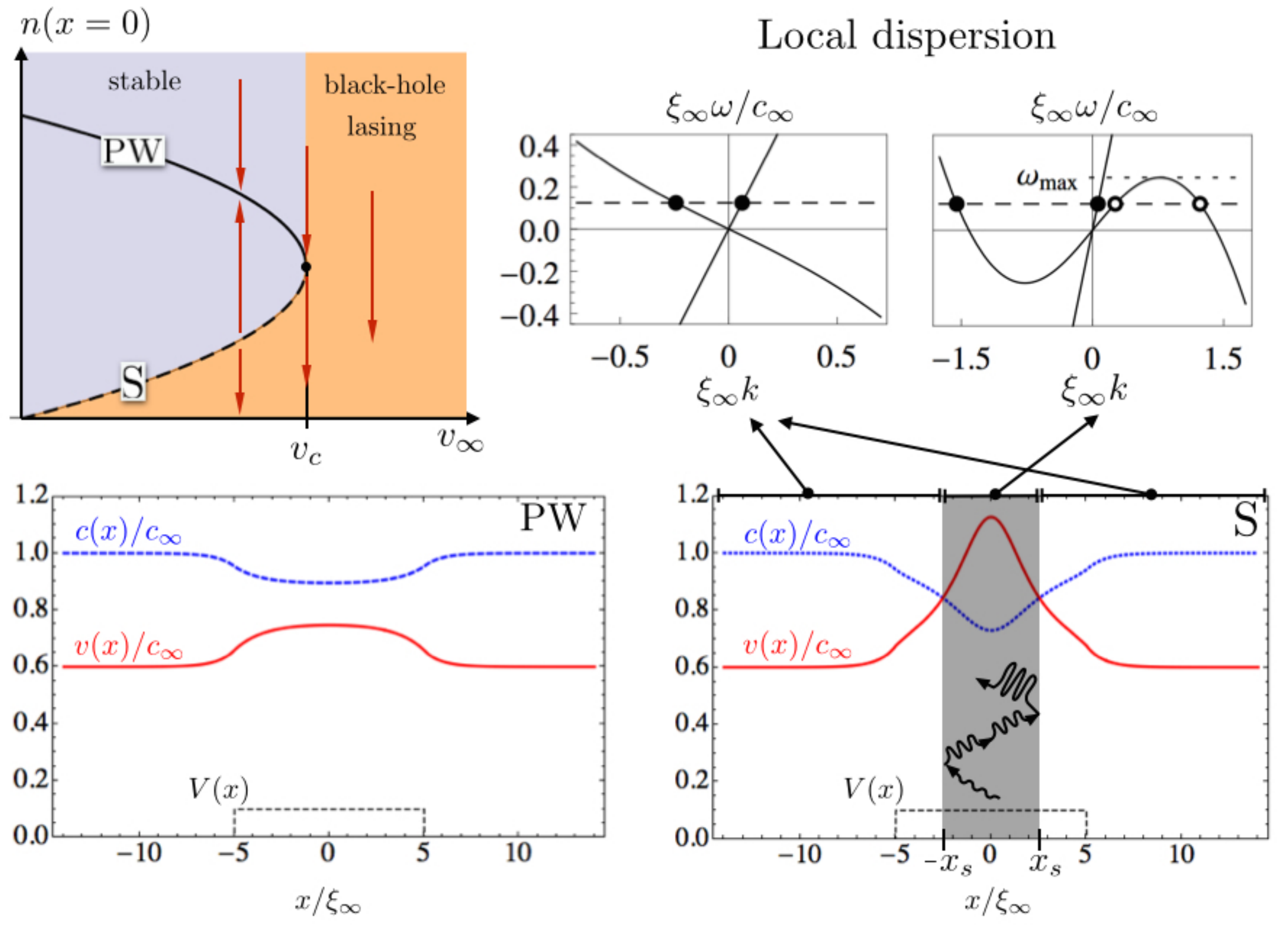}
\caption{Upper left: dynamical phase diagram of the saddle-node bifurcation charecterizing the GP equation \eqref{eq:GP} at fixed injected current. On the vertical axis the density at the barrier center is reported, distinguishing the PW solution from the other stationary solution S. The arrows indicate the direction of the dynamical evolution, separating a stable region converging toward the PW attractor from an unstable region, delimited by the unstable S solution, where no convergence is present. Lower row: local flow velocity (red solid line) and sound speed (blue dotted line) for the PW and S solutions at a given injected velocity $v=0.93v_c$, barrier height $V_0=0.1gn_\infty$ and width $d=5\xi_\infty$, with $\xi_\infty$ being the healing length far away from the barrier. Upper right panels: local dispersion relation $(\om-v(x)k)^2=c(x)^2k^2+\frac{\hbar^2k^4}{4m^2}$ of small amplitude modes in the subsonic (left) and supersonic (right) region. The additional modes appearing in the latter propagate back and forth and get amplified, giving rise to the black-hole lasing dynamical instability.}
\label{fig:bifurcation}
\end{figure}

Focusing on the one-dimensional (1D) flow of a BEC through a penetrable repulsive barrier, we show that the instability governing the underlying saddle-node bifurcation \cite{hakim,brachet_2002} is actually a dynamical black-hole lasing effect, triggered by a finite set of propagating modes which resonate within an effective cavity formed by two horizons enclosing the obstacle. The position of these sonic horizons is set by $v(x)=c(x)$ (see Fig.~\ref{fig:bifurcation}), with $v(x),c(x)$ being the local flow velocity and sound speed, respectively. Even for a barrier much thinner than the BEC coherence length, the above local quantities are of physical relevance, providing additional negative energy modes inside the cavity where $v(x)>c(x)$.
We analytically show that the critical velocity and decay rate depend directly on the horizons configuration, and only indirectly on the shape of the barrier and interaction strength, i.e., two very different barriers can give rise to a comparable critical velocity and decay rate, offering a means of experimental testing. 
For obstacles much broader than the coherence length, one recovers the continuous spectrum of negative energy modes at the basis of Hawking radiation from a single horizon and of the Landau energetic instability, which is triggered by the presence of an impurity in homogeneous supersonic flows.

\section{Model setup and saddle-node bifurcation}

A BEC flowing in one direction with a strong transverse confinement can be well described by the 1D GP equation \cite{pit_str_book}:
\begin{equation}
\label{eq:GP}
i\hbar\partial_t \psi(x,t)=-\left(\frac{\hbar^2}{2m}\partial_{xx}+V(x)+g|\psi(x,t)|^2\right)\psi(x,t)\;,
\end{equation}
where $\psi(x,t)=\sqrt{n(x,t)}e^{i\phi(x,t)}$ is the superfluid order parameter (or the BEC wavefunction), $V(x)$ is a repulsive square barrier potential of width $2d$ and height $V_0$, and $g$ is the effective 1D interaction strength. We model the flow by imposing that the density $n=|\psi|^2$ and velocity $v=\hbar\partial_x\phi/m$ take constant values $n_\infty$, $v_\infty$, respectively, far from the barrier at $|x|\to\pm\infty$. This is completely equivalent to solving the GP in a moving frame without the above boundary conditions, which in turn describes the case of a moving barrier in a standing BEC \cite{hakim}. 
The solutions of the nonlinear Eq.~\eqref{eq:GP} with these boundary conditions show a saddle-node bifurcation \cite{baratoff,sols,hakim,brachet_2002,piazza_cp} at a critical velocity $v_c$ (or barrier height $V_c$), where the only two stationary solutions merge and disappear. These two solutions, shown in the lower panels of Fig.~\ref{fig:bifurcation}, become a plane wave and a soliton when the height of the barrier goes to zero. For this reasons, in what follows we shall refer to them as the PW solution and the S solution. As indicated by the arrows in the upper left panel of Fig.~\ref{fig:bifurcation}, the PW solution is a stable attractor within the parameter region delimited by the S solution. As verified numerically with GP dynamics \cite{hakim} and linear stability analysis \cite{brachet_2002,kato_prl_2010}, the latter is instead dynamically unstable. This unstable behavior, characterized by soliton emission, is also present in the whole region above $v_c$, where the emitted solitons belong to a nonlinear dispersive shockwave \cite{gur_pitav_1974,hakim,lesz}. The dynamical saddle-node phase diagram of the upper left panel in Fig.~\ref{fig:bifurcation} can be modeled by the equation $\dot{f}(t)=(v_c-v)-f^2(t)$, where the unstable region corresponds to a function $f$ diverging in time $t$. This simple model also explains the universality of the dynamics on all sides of the the bifurcation, which has been verified numerically in \cite{brachet_2002}.

\section{The black-hole lasing effect}

In the following, we show that the dynamical instability characterizing the GP saddle-node bifurcation, and thereby responsible for the superflow decay, is due to the black-hole lasing effect, whose origin is the presence of a black-hole-white-hole pair of sonic horizons. 
To this purpose, we study perturbations on top of the stationary solutions by means of the Bogoliubov--de Gennes equation.
The latter can be written in terms of the flow velocity $v(x)$ and local speed of sound $c(x) =\sqrt{gn(x)/m}$ only~\cite{MacherBEC}.
Therefore, the excitation spectrum and the presence and the nature of any instability are fully governed by $v(x)$ and $c(x)$. 
Remarkably, this property, which allows for the mapping to the Klein--Gordon equation describing the propagation of a scalar field on a curved spacetime~\cite{livrev}, holds even when the local density description of the flow breaks down, i.e., even when the healing length $\xi(x)=\hbar/2mc(x)$ exceeds the barrier width.

As it appears from the lower right panel of Fig.~\ref{fig:bifurcation}, the S solution always presents two sonic horizons at $x=\pm x_s$, delimiting a compact supersonic region where the local flow velocity exceeds the sound speed.
The point $-x_s<0$ such that $v(-x_s)=c(-x_s)$ behaves thus as the analog of a black hole horizon, in the sense that phonons cannot propagate from the internal to the external region.
Similarly, the point $+x_s$ such that $v(x_s)=c(x_s)$ is the analog of a white hole horizon.
In particular, when the distance between the horizons is large enough, the spectrum of perturbations is enriched by additional propagating modes characterized by a negative norm, thus carrying negative energy as seen by the laboratory reference frame.
As shown in Fig.~\ref{fig:bifurcation}, the Bogoliubov dispersion relation 
\begin{equation}\label{eq:disp}
(\om-vk)^2=c^2k^2+\frac{\hbar^2k^4}{4m^2}
\end{equation}
admits indeed only two real solutions of the wavenumber $k$ for each frequency $\omega$ in the subsonic region, corresponding to two standard left and rightward propagating modes. Instead, in the supersonic region it has four real solutions for $\om$ smaller than a certain threshold frequency $\ommax$, two of them (open dots) lying on the negative-norm branch of the dispersion relation.
This bounded anomalous modes give rise to anomalous transmission and reflection, ultimately leading to a cavity amplification effect generating a dynamical instability known as black-hole laser effect~\cite{cj,carlos,ulf,bhlaserrevisited,piyush,coutant,bhlaserBEC}.
More precisely, Hawking-like phonons are emitted by the analog horizons $x=\pm x_s$~\cite{garay} due to the anomalous scattering of negative- and positive-norm modes. These horizons act as amplifiers of phonons~\cite{unruhamplifier}, converting a negative-norm wave of unitary amplitude into two negative- and positive-norm waves of amplitudes $\alpha$ and $\beta$, respectively, with $|\alpha|^2-|\beta|^2=1$ and $|\alpha|>1$. Since the supersonic region is compact, the negative-norm waves, one leftgoing and one rightgoing, are trapped and can be interpreted as a single mode bouncing back and forth between the two sonic points. Consequently, the internal region acts as a resonant cavity for this mode which exponentially grows being amplified at any bounce on the sonic horizons.
Formally, this leads to the appearance of positive imaginary parts in the Bogoliubov frequency spectrum.

However, when the horizons are close to each other, a naive analysis of the dispersion relation Eq.~(\ref{eq:disp}) is not sufficient to properly describe the properties of the spectrum of these modes. As shown below, one can derive a generalised Bohr-Sommerfeld quantization condition [see Eq.~\eqref{eq:instability}] which fixes the number of resonant negative-norm cavity modes, their frequencies, and growing rates.
Accordingly, negative energy modes can appear only if at least one oscillation can be hosted within the two classical turning points $\pm x_s$ of the Bogoliubov--de Gennes equation. Thus, the presence of a region with $c(x)\leq v(x)$ is necessary but not sufficient for existence of negative energy modes. Remarkably, the appearance of negative energy modes in a compact supersonic region has been experimentally verified using Bragg spectroscopy~\cite{steinhauer_2010}.  

In the following, we provide the main results of our calculations, which are presented in detail in the Supplemental Material (SM).
We generalise the methods developed in \cite{bhlaserBEC,coutant,parentanimichel,broad_hor,coutantbroadhor} for steplike flow profiles with constant velocity to the physically relevant configuration (see Fig.~\ref{fig:bifurcation}) discussed in the present work.
It can be shown that, when the fluid velocity is increased, the first unstable mode appears at $\omega=0$. When the flow is symmetric with respect to $x=0$ (more general cases are considered in the SM), this implies that the flow is unstable when
\begin{equation}\label{eq:instability}
\frac{2m}{\hbar}\int_{-x_s}^{x_s}dx\sqrt{v(x)^2-c(x)^2}\geq\arg{\frac{\beta}{\alpha}}.
\end{equation}
The left-hand side of this inequality is the phase acquired by a zero-frequency mode propagating across the supersonic region. The right hand side is instead associated to the phase acquired by the mode when scattered at the sonic point.

\begin{figure}
\includegraphics[width=12.0cm]{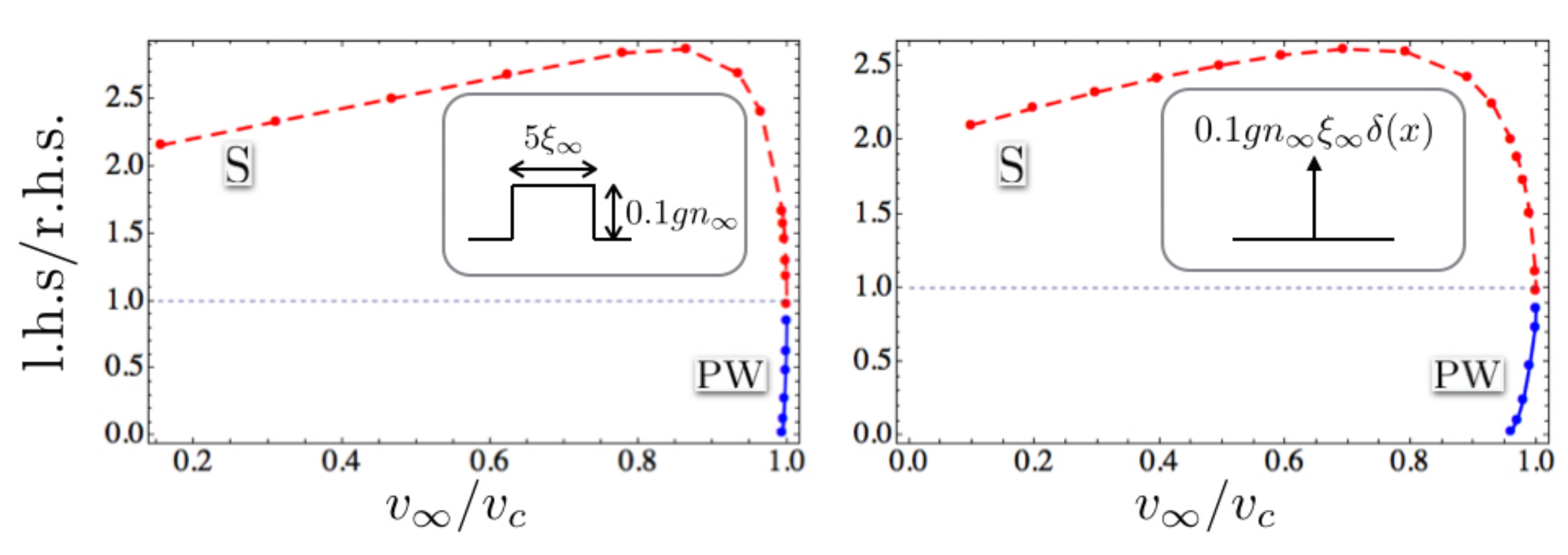}
\caption{\label{fig:lasingcondition}Ratio between the left and the right hand sides of  Eq.~\eqref{eq:instability} for the PW (blue dots with solid line) and S (red dots with dashed line) solutions in the hydrodynamic regime (left panel, parameters as in the lower panels of Fig.~\ref{fig:bifurcation}) and for the delta barrier potential (right panels, $V=0.1gn_\infty\xi_\infty\,\delta(x)$).}
\end{figure}
The ratio of the two sides of this inequality is reported in Fig.~\ref{fig:lasingcondition} for the PW (closed dots) and S solutions (open dots) in the hydrodynamic regime (left panel) and for a delta barrier potential (right panel). This ratio is always smaller than one for the PW solution, which therefore never supports a cavity mode, while it is greater then one for the S solution, which then always supports a cavity mode. The main result is that at the bifurcation point both solutions reach the marginal condition for the appearance of the cavity mode. 
Moreover, the condition Eq.~\eqref{eq:instability} allows for an analytical prediction of the critical velocity $v_c$ corresponding to the bifurcation point. In the case where the fluid velocity and the speed of sound assume step like profiles, i.e., $v_\infty=v(|x|>x_s)$, $v_{barr}=v(|x|<x_s)$ and the same for $c(x)$, the critical injected velocity $v_c$ is obtain by solving
\begin{equation}
\frac{2m x_s}{\hbar}\sqrt{v_{barr}^2-c_{barr}^2}=\arctan\frac{\sqrt{v_{barr}^2-c_{barr}^2}\sqrt{c_\infty^2-v_\infty^2}}{v_{barr} v_\infty-\lambda\, c_{barr}^2},
\end{equation}
for $v_\infty$, and $\lambda=[1+\ln(v_\infty/v_{barr})]/[1-\ln(v_\infty/v_{barr})]$.
Note that the resulting critical velocity depends on the barrier potential only through the local sound speed and flow velocity, accordingly to the adopted analog gravity description. The most general formula valid for any smooth profile is derived in the SM.

\begin{figure}
\includegraphics[width=12cm]{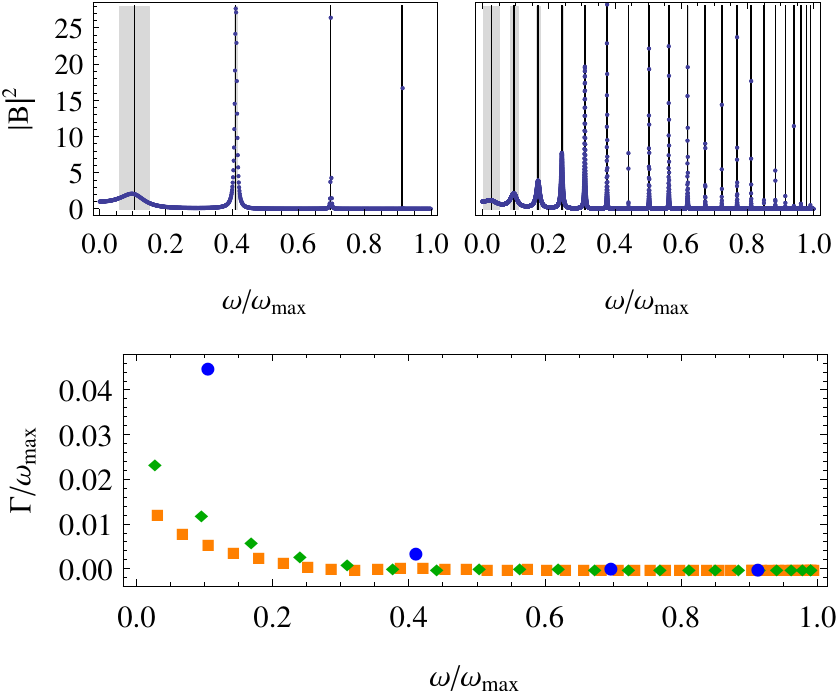}
\caption{\label{fig:spectrum} (color online) Upper panels: Squared modulus of the amplitude $B$ of the cavity mode (blue dots) for a unitary incoming leftgoing wave for two values of the distance $2$ between the sonic points ($2x_s/\xi_s=10$, left; $2x_s/\xi_s=50$, right). Vertical lines denote the real parts of the complex eigenfrequencies $\om_n+i\Gamma_n$ and the shaded areas represent the intervals $(\om_n-\Gamma_n,\om_n+\Gamma_n)$. Lower panel: Spectra for $2x_s/\xi_s=10,\,50,\,100$ (blue dots, green diamonds, orange squares). For $2x_s\to\infty$, $\Gamma_n\to0$ and the spectrum becomes dense in the real interval $(0,\ommax)$.}
\end{figure}

As anticipated, the anomalous cavity modes are always dynamically unstable, having positive imaginary parts $\Gamma_n$ of the complex eigenfrequencies.
Typical results, computed using the algorithm developed in Ref.~\cite{bhlaserBEC} are presented in Fig.~\ref{fig:spectrum}.
In the upper panels, the blue dots represent the value of the cavity mode amplitude $|B|^2$ as a function of energy $\om$ and for two different cavity lengths $2x_s$.

In the lower panel the imaginary part is shown. As the cavity width $2x_s$ grows, the number of eigenfrequencies increases while their imaginary parts vanish, such that the spectrum becomes dense in the real interval $(0,\ommax)$.
Thus, for $x_s\to\infty$, phonons are spontaneously emitted for all frequencies $\om<\ommax$, the emission rate is constant in time ($\Gamma_n=0$), and the instability is energetic rather than dynamical. In this limit two different mechanisms can excite the same continuous set of negative-energy modes: (i) Hawking-like pairs of phonons are emitted from each of the two far apart sonic horizons~\cite{MacherBEC}; (ii) In the presence of impurities phonons are produced by Landau instability. In this regime, achievable with a broad potential such that $2x_s\gg \xi_s$, the critical condition corresponds to the flow velocity reaching the sound speed inside the barrier region, which in our one-dimensional configuration means  $v(x=0)=c(x=0)$. This criterion has been numerically verified in the hydrodynamic regime of GP equation in one \cite{watanabe}, two \cite{rica,adams,mathey,piazza_2009} and three dimensions \cite{piazza_2011,piazza_2013,feder} and also for a fermionic superfluid using the Bogoliubov-de Gennes equations \cite{spunt}, as well as in the toroidal BEC experiments of \cite{hadzibabic_2012,nist_criticalvelocity_2013}.

\section{Conclusions}

By showing that the instability of the one-dimensional BEC flow through a penetrable barrier is due to the dynamical black-hole lasing effect, this work builds a bridge between field theory in curved spacetime and superfluidity. It provides a deeper insight into the long standing problem of supercurrent instability and also identifies an experimental available setup as a natural candidate to observe interesting physics of fluctuations in curved spacetime.

In particular, we have shown that the system's behavior is governed by the sonic horizons configuration and the (mis)match in the number of excitation modes on the two sides of them. This allowed us to provide a general formula predicting the critical velocity and to characterize the crossover between the dynamical black-hole lasing instability and the Hawking energetic instability. The latter coincides with the well-known Landau instability and is achieved for broad enough barrier potentials.

The present analysis allows also for a suggestive explanation of the known supercritical stationary flow, found for velocities $v_\infty$ larger than a second higher critical velocity \cite{pavloff_2002,engels_2007,lesz}.
For this stationary configuration, the flow around the barrier is slower than outside.
If $v_\infty$ is large enough the flow will be everywhere supersonic~\cite{lesz}. In this case, at low frequencies, no mode mismatch is present, since there are four propagating modes both in the internal and the external regions~\cite{supersonic}.
If $v_\infty$ is maintained supersonic but decreased below this higher critical velocity, the internal region may become subsonic. In this case, mode mismatch reappears and phonons are expected to be emitted with a linearly growing rate, as suggested by studies in the context of Lorentz violating quantum field theories~\cite{warpdriveBEC}.

Our study does not include the possiblity for fluctuations (thermal or quantum) to trigger the instability of the PW branch before the critical saddle-node bifurcation point is reached. However, the black-hole lasing instability should be the relevant decay mechanism also in this case. It has been indeed shown \cite{levine_1997} that fluctuations lead the stable PW branch to tunnel into the unstable S solution. It is the dynamical instability of the latter, which we have shown to be due to black-hole lasing, that leads to soliton emission. 

An extension of this study to the two- or three-dimensional flow, where a more complicated horizon configuration appears, would be the object of future study.

\vspace{1cm}

\emph{Note added.}
Experimental evidence of black hole laser effect has been very recently reported by Jeff Steinhauer in Ref.~\cite{steinhauerobs}.

\begin{acknowledgments}

We are grateful to N. Pavloff and  J. Steinhauer for very useful comments.
We thank I. Carusotto, O. Gat for stimulating discussions.
This work has been financially supported by ERC through the QGBE grant and by Provincia Autonoma di Trento.
A. R. and F. P. acknowledge support from the Alexander Von Humboldt foundation.

\end{acknowledgments}

\bibliography{literature}

\section{Supplemental Material}

\subsection{Scattering matrix}

We perform a linear stability analysis of a one-dimensional flowing condensate, whose superfluid velocity and speed of sound are shown in Fig.~1 of the Main Text (lower right panel), with two sonic horizons at $x=\pm x_s$, such that the flow is supersonic in the compact region $-x_s<x<{x_s}$ and subsonic elsewhere. The point $x={x_s}$ behaves as the analog of a black hole horizon, in the sense that phonons cannot propagate from the internal to the external region, in much the same way as light cannot escape from the interior of a black hole. Similarly, the point $x=-{x_s}$ is the analog of a white hole horizon.

Adopting the scattering matrix approach of~\cite{coutant}, we neglect the scattering between forward and contra-propagating modes (with respect to the fluid velocity) and consider only contra-propagating modes. Thus, the scattering process at both the horizons $x=\pm {x_s}$ can be described in term of two modes. One of them, with positive norm, can cross the horizons from right to left and corresponds to the leftmost solution of the dispersion relation in the upper central (subsonic flow) and right (supersonic flow) panels of Fig.~1. The second one, with negative norm, bounces back and forth between the horizons and corresponds to the two open-dot solutions of the upper right panel of Fig.~1, representing, respectively, the ingoing and outgoing branches of the mode for the left horizon and vice versa for the right horizon.

The complex unstable modes are constructed by fixing to zero the amplitude of the incoming positive norm mode entering in the system for the right subsonic asymptotic region, and imposing that the internal bouncing negative norm mode assumes the same values after a full bounce process from one horizon to the other and back to the first one. This gives a sort of quantization condition, which, as long as the wavelength of the modes is short with respect to the scale of variation of the fluid velocity and speed of sounds, can be expressed through the WKB approximations
\begin{equation}\label{eq:S22}
\tilde\alpha_B\tilde\alpha_W e^{-i(S_1-S_2)}\left(1+\frac{\tilde\beta_{B}}{\tilde\alpha_B}\frac{\beta_W}{\tilde\alpha_W}e^{i(S_++S_1)}\right)=1,
\end{equation}
where
\begin{equation}\label{eq:S}
S_+=-\int_{-{x_s}}^{{x_s}}dx\,k_+,\quad S_{1,2}=\int_{L_{\om}}^{R_{\om}}dx\,k_{1,2}
\end{equation}
are the phase acquired by the positive norm mode with wavenumber $k_{+}$ (closed dot, leftmost solution of the dispersion relation in the upper right panel of Fig.~1) in the propagation from the white hole ($x={x_s}$) to the black-hole ($x=-{x_s}$) and by the two negative norm modes with wavenumbers $k_{1,2}$ (open dots, $k_1$ is the largest of the two solutions) in the propagation between $L_{\om}$ and $R_{\om}$, the turning points of the two associated modes~\footnote{To be consistent we changed sign to the corresponding expressions in~\cite{coutant,bhlaserBEC}, where the fluid propagated with $v<0$, the white horizon was located at $x=-{x_s}$ and the black one at $x=+{x_s}$}.
\begin{equation}
U_{B}=
\begin{pmatrix}
\alpha_{B} & \beta_{B}\\
\tilde\beta_{B} & \tilde\alpha_{B}\\
\end{pmatrix},\qquad
U_{W}=
\begin{pmatrix}
\alpha_{W} & \beta_{W}\\
\tilde\beta_{W} & \tilde\alpha_{W}\\
\end{pmatrix}
\end{equation}
are the scattering matrix connecting the incoming and outgoing field $(\phi_+,\phi_-)$ with positive and negative norm components $\phi_+$ and $\phi_-$, satisfying
\begin{equation}\label{eq:unitarity}
U\eta U^\dagger=\eta,\qquad
\eta=\begin{pmatrix}
1 & 0\\
0 & -1
\end{pmatrix}.
\end{equation}
This implies
\begin{equation}\label{eq:norms}
|\alpha|^2-|\beta|^2=1,\quad |\tilde\alpha|^2-|\tilde\beta|^2=1,\quad\frac{\beta}{\alpha}=\left(\frac{\tilde\beta}{\tilde\alpha}\right)^*.
\end{equation}
Furthermore, since a white hole is the time reversal of a black hole, the scattering matrix of a white hole can be computed by exchanging ingoing and outgoing modes (i.e., taking the inverse of the scattering matrix) and by taking the complex conjugate of all modes and coefficients:
\begin{equation}
U_W=(U_B'^{-1})^*=\eta U_B'^T \eta,
\end{equation}
where we have used Eq.~\eqref{eq:unitarity}, $^T$ denotes the transpose operator, and we have denoted by $U_B'$ the scattering matrix of the corresponding black hole, to distinguish it from the scattering matrix $U_B$ of the black hole located at $x=-{x_s}$. For later convenience we explicitly write the relation among the coefficients of $U_W$ and $U_B'$:
\begin{equation}\label{eq:WB}
\alpha_W=\alpha_B',\quad\tilde\alpha_W=\tilde\alpha_B',\quad\beta_W=-\tilde\beta_B',\quad\tilde\beta_W=-\beta_B'.
\end{equation}
%

\subsection{High frequency}

No assumption has been made yet on the form of the matrices $U_{B,W}$. In the high frequency limit the ratios $\beta/\alpha$ exponentially vanish~\cite{birreldavies} and the term in parenthesis on the left-hand side of  Eq.~\eqref{eq:S22} is 1. Following Ref.~\cite{coutant}, the real part $\om_n$ of their complex eigenfrequencies are fixed by a Bohr-Sommerfeld condition for integer $n\geq1$,
\begin{equation}\label{eq:bs}
 \int_{L_{\om_n}}^{R_{\om_n}}\!\!\!dx\left[-k_{\om_n}^{(1)}(x)+k_{\om_n}^{(2)}(x)\right]
 +\phi_{B}(\om_n)+\phi_{W}(\om_n)=2\pi n,
\end{equation}
where $\phi_{B,W}=-\arg(\tilde\alpha_{B,W})$ are the phases acquired by the negative frequency mode when bouncing on the black and white horizons, respectively. A more refined calculation~\cite{coutant,bhlaserBEC} allows one to compute also the positive imaginary parts $\Gamma_n$ of the complex eigenfrequencies, corresponding to modes exponentially growing as $e^{\Gamma_n t}$.
To determine the complex eigenfrequencies $\om_n+i\Gamma_n$, which are reported in Fig.~3 of the Main Text, one can exploit the fact that the squared amplitude $|B(\om)|^2$ of the cavity mode when the system is stimulated by a leftgoing plane wave of unitary amplitude and real frequency $\om$ is a sum of Lorentzians of width $\Gamma_n$ centered at $\om_n$~\cite{coutant}.

\subsection{Low frequency}
While the technique presented in the above section allows to follow the evolution of the modes by varying the parameters of the systems, it is not suitable to describe the birth of new unstable modes when increasing ${x_s}$. In particular it does not allow to estimate under which conditions the first complex mode appears, that is when the system starts to be unstable. In this case, the opposite limit $\om\to0$ must be considered.

However, in this limit one must carefully check that Eq.~\eqref{eq:S22} is still valid. In this limit, using the dispersion relation of Eq.~2 of the Main Text,
\begin{equation}
k_1(x)=-k_+(x)=\frac{2m}{\hbar}\sqrt{v(x)^2-c(x)^2},
\end{equation}
which is large (order of $1/\xi$, where $\xi$ is the healing length of the condensate) far enough from the horizon, so that these modes can be described in the WKB approximations. Moreover, in this limit $R_{\om->0}=x_s$ and $L_{\om->0}=-x_s$, so that Eq.~\eqref{eq:S} yields
\begin{equation}
S_1=S+=\frac{2m}{\hbar}\int_{-{x_s}}^{{x_s}}dx\sqrt{v(x)^2-c(x)^2}.
\end{equation}
Conversely, $k_2=0$ for $\om=0$, so the wavelength is infinite and the WKB approximation is no longer valid. However, in this case, the contribution $S_-$ of this mode to the total phase vanishes, and Eq.~\eqref{eq:S22} is still valid. Using Eq.~\eqref{eq:WB}, Eq.~\eqref{eq:S22} can then be written as
\begin{equation}
\tilde\alpha_B\tilde\alpha_B' e^{-iS_+}-\tilde\beta_{B}\tilde\beta_B'e^{iS_+}=1.
\end{equation}

Furthermore, for $\om\to0$ the coefficients of the scattering matrices diverge~\cite{birreldavies} and the normalization constraints~\eqref{eq:norms} implies that $|\tilde\beta|\to|\tilde\alpha|$. Consequently,
\begin{equation}
e^{-iS_++i\arg{\tilde\alpha_B}+i\arg{\tilde\alpha_B'}}-e^{iS_++i\arg{\tilde\beta_B}+i\arg{\tilde\alpha_B'}}
=\frac{1}{|\tilde\alpha_B\tilde\alpha_B'|}.
\end{equation}
Since the right-hand side vanishes for $\om\to0$, the exponents at left-hand side must be equal modulus $2n\pi$:
\begin{equation}
2S_+=\arg{\frac{\tilde\alpha_B}{\tilde\beta_B}}+\arg{\frac{\tilde\alpha_B'}{\tilde\beta_B'}}+2n\pi.
\end{equation}
Finally, using the last equation of Eq.~\eqref{eq:norms},
\begin{equation}\label{eq:instcondition}
S_+=\frac{1}{2}\left(\arg{\frac{\beta_B}{\alpha_B}}+\arg{\frac{\beta_B'}{\alpha_B'}}\right)+n\pi.
\end{equation}
%

\subsubsection{Step-like profiles with homogeneous velocity}

In the case of a step-like profile and for a homogeneous flow velocity, the expression $\beta/\alpha$ in the right-hand side can be computed for each horizons using the results of~\cite{mayoral}
\begin{equation}\label{eq:ratiocarlos}
\frac{\beta}{\alpha}=\frac{\sqrt{v_0^2-c_p^2}+i\sqrt{c_b^2-v_0^2}}
{-\sqrt{v_0^2-c_p^2}+i\sqrt{c_b^2-v_0^2}},
\end{equation}
and
\begin{equation}
\arg\frac{\beta}{\alpha}=2\arctan{\sqrt{\frac{c_b^2-v_0^2}{v_0^2-c_p^2}}}
\end{equation}
Since both this quantity and $S_+$ are positive, the first unstable mode appears for $n=0$. When the white hole and the black hole are symmetric, that is $\beta_B=\beta_B'=\beta$ and $\alpha_B=\alpha_B'=\alpha$, this first mode appear at ${x_s}=x_0$, satisfying
\begin{equation}\label{eq:renaud}
x_0=\frac{\hbar}{2m\sqrt{v_0^2-c_p^2}}\arctan\sqrt\frac{c_b^2-v_0}{v_0^2-c_p^2}.
\end{equation}
This equation agrees with the results of Ref.~\cite{parentanimichel}, where the condition for the appearance of new modes was for the first time analytically computed in this configuration.

\subsubsection{Step-like profiles with inhomogeneous velocity}

In the more general case, where the velocity profile is steplike, but the velocity of the fluid assumes different values $v_p$ and $v_b$ in the internal supersonic region and in the external subsonic one, respectively, the general matching conditions for the phonon field where found in~\cite{tworegimes}. Starting from the Bogoliubov--de Gennes equation
\begin{equation}\label{eq:bdg}
  i\pdt\hp = \left[T_v-i v \pdx + m c^2 \right] \hp + m c^2 \hpd,
\end{equation}
for the evolution of the phonon field on a top of a flowing BEC, where
\begin{equation}
 T_v=-\frac{ 1 }{2m}\pdx^2 +\frac{ 1 }{2m} \left[\pdx\log(v(x))\right]\pdx
\end{equation}
is a differential operator acting on $\hp$, and assuming a step-like velocity profile of the form
\begin{align}
 v(x)&=v_b\,\theta(x_s-x)+v_p\,\theta(x-x_s),\\
 c(x)&=c_b\,\theta(x_s-x)+c_p\,\theta(x-x_s),
\end{align}
where $x_s$ is the location of the sonic point, the solution of Eq.~\eqref{eq:bdg} can be written as a sum of plane waves in both the homogeneous regions on the two sides of horizon. To obtain globally defined solutions one must derive the matching conditions for $\hp$ and its first spatial derivative at $x=x_s$, by integrating Eq.~\eqref{eq:bdg} in a neighborhood of $x_s$. Following~\cite{tworegimes} one finds that $\phi$ is continuous across the horizon while its first derivative undergoes a jump:
\begin{equation}
 \pdx\hp_b|_{x\to x_s^-}=\lambda\,\pdx\hp_p|_{x\to x_s^+},
\end{equation}
where
\begin{equation}
\lambda=\frac{1+\ln(v_b/v_p)}{1-\ln(v_b/v_d)}
\end{equation}

The generalization of Eq.~\eqref{eq:ratiocarlos} is obtained after some lengthy algebra
\begin{equation}
\frac{\beta}{\alpha}=\frac{(v_p v_b-\lambda\, c_p^2)+i\sqrt{v_p^2-c_p^2}\sqrt{c_b^2-v_b^2}}
{-(v_p v_b-\lambda\, c_p^2)+i\sqrt{v_p^2-c_p^2}\sqrt{c_b^2-v_b^2}},
\end{equation}
from which
\begin{equation}\label{eq:betaoveralpha}
\arg\frac\beta\alpha=2\arctan\frac{\sqrt{v_p^2-c_p^2}\sqrt{c_b^2-v_b^2}}{v_p v_b-\lambda\, c_p^2}.
\end{equation}
Inserting Eq.~\eqref{eq:betaoveralpha} in Eq.~\eqref{eq:instcondition}, one obtains Eq.~(4) (where $v_{b,p}=v_{\infty,{\rm barr}}$, $c_{b,p}=c_{\infty,{\rm barr}}$), which is valid for general step-like profiles with arbitrary fluid velocities and speeds of sound in the three regions defined by the two horizons.

\subsubsection{Continuous profiles}
Unfortunately, it is very difficult to realize such a double-step profile in realistic experimental conditions. In order to investigate the experimental feasibility of this system and the possibility of observing the lasing effect, the above conditions must be generalized to continuous transitions between subsonic and supersonic region. To this aim, we use the results of Ref.~\cite{broad_hor} (recently confirmed by~\cite{coutantbroadhor}), where it was noticed that the temperature of the Hawking emission can be computed using an effective surface gravity $\bar\kappa$, given by the spatial average
\begin{equation}
 \bar\kappa = \frac{1}{\Delta x} \int_{x_s-\Delta x/2}^{x_s+\Delta x/2} dx\, \kappa(x)
\end{equation}
of the local surface gravity
\begin{equation}
 \kappa(x)=\frac{d\left[c(x)-|v(x)|\right]}{dx}
\end{equation}
over a broad region center around $x_s$ of width
\begin{equation}\label{eq:kav}
 \Delta x=\xi_s\left(\frac{2 c_s}{\kappa\,\xi_s}\right)^{1/3},
\end{equation}
where $c_s$ and $\xi_s=\hbar/\sqrt{2}mc_s$ and are the speed of sound and the healing length, respectively, calculated at the horizon and $\kappa$ is the surface gravity at the horizon. Furthermore, in~\cite{tworegimes}, it was verified with numerical simulations that the temperature computed with this effective surface gravity is a very good approximation for the temperature of the emission for any kind of horizons, from very smooth transition (standard Hawking regime) to step-like transitions.

In view of those results, in this work we have assumed that, for any given smooth profile defined by its fluid velocity $v(x)$ and by its speed of sound $c(x)$, there exists an equivalence class of profiles with the same scattering matrix. In particular, in this class one can always identify a step like profile
\begin{align}
 &v_p = v(x_s-\Delta/2), &v_b = v(x_s+\Delta/2),\\
 &c_p = c(x_s-\Delta/2), &c_b = c(x_s+\Delta/2),
\end{align}
obtained by computing $v$ and $c$ at a distance $\pm \Delta/2$ from the horizon. This effective step-like profile can then be used to evaluate the sought-after expression of Eq.~\eqref{eq:betaoveralpha} for the corresponding smooth velocity profile. The results reported in Fig.~2 of the Main Text have been obtained by this technique.

\end{document}